# DEGREES OF FREEDOM OF TONGUE MOVEMENTS IN SPEECH MAY BE CONSTRAINED BY BIOMECHANICS

*Pascal Perrier[1], Joseph Perkell[2], Yohan Payan[3], Majid Zandipour[2], Frank Guenther[4] and Ali Khalighi[1]*

[1]Institut de la Communication Parlée, UMR CNRS 5009– INPG & Université Stendhal - Grenoble – France

[2.]Speech Communication Group – RLE – MIT – Cambridge – Massachusetts – USA

[3]TIMC – Université Joseph Fourier – Grenoble – France

[4]CNS-CAS, Boston University, Boston, Massachusetts, USA

Contact : perrier@icp.inpg.fr

## ABSTRACT

A number of studies carried out on different languages have found that tongue movements in speech are made along two primary degrees of freedom (d.f.s): the high-front to low-back axis and the high-back to low-front axis. We explore the hypothesis that these two main d.f.s could find their origins in the physical properties of the vocal tract. A large set of tongue shapes was generated with a biomechanical tongue model using a Monte-Carlo method to thoroughly sample the muscle control space. The resulting shapes were analyzed with PCA. The first two factors explain 84% of the variance, and they are similar to the two experimentally observed d.f.s. This finding suggests that the d.f.s are not speech-specific, and that speech takes advantage of biomechanically based tongue properties to form different sounds.

## 1. INTRODUCTION

The production of speech requires the simultaneous control of more than thirty different muscles. At the same time, the classical articulatory description of vowel production is based on a small number of parameters: high/low, front/back, rounded/spread. Hence the understanding of speech motor control strategies and the construction of models for this control would seem to require a reduction of the dimensionality from the muscle control space to a more functional, speech-related control space. The dimensions of the functional control space are then called the degrees of freedom of the vocal tract.

The desired features for such a dimensionality reduction are (1) a capability for generalization across speakers of a particular language in order to provide information primarily about the articulation of speech in this language and not about individual speaker characteristics, and (2) an interpretability in terms of muscles synergies and antagonisms in order to provide an understandable view of the way muscles are actually coordinated to produce speech. In addition, comparing low-dimensional control spaces across languages (dimensionality, directions of the degrees of freedom) and the relations between theses functional spaces and the muscle control space provides a interesting way to quantitatively compare the production of different languages.

A number of studies have dealt in the past years with the issue of dimensionality reduction. These were all based on statistical analyses of articulatory data. Harshman et al. (1977), Jackson (1988), Nix et al. (1996) and Hoole (1998) applied a PARAFAC analysis to Xray or EMA data for English, Icelandic or German. Maeda (1990) used a guided principal component analysis (PCA) for Xray data on French. Sanguineti and colleagues (Sanguineti et al., 1997; Sanguineti et al., 1998) used the same corpus as Maeda, but used a biomechanical model of the tongue, jaw and hyoid bone to provide a projection of the data set in a modeled muscle control space, in which they ran a PCA. These authors were the first to provide not only a reduction of dimensionality, but also a description of the muscular correlates of the degrees of freedom of the articulations.

It is interesting to observe that while four languages were analyzed in these different studies, most of them have found that more than 90% of the variance observed in tongue shapes during speech can be described along two primary degrees of freedom: (1) movement of the tongue body along a high-front to low-back axis (called *front raising* in Harshman et al., 1977) and (2) bunching of the tongue along a high-back to low-front axis (called *back raising* in Harshman et al., 1977). Jackson (1988) found that the number of degrees of freedom were different for English and Icelandic, but his PARAFAC analysis was then proved to be degenerate by Nix et al. (1996), who reanalyzed the same data sets. These results lead to a question about the origin of these two main degrees of freedom: are they learned, speech-specific actions; are they in some way basic properties of the production mechanism; or are they due to a combination of influences?

In an EMG study of one subject, Maeda and Honda (1996) found that combinations of the hyoglossus, styloglossus and different parts of the genioglossus muscle act as agonist-antagonist pairs to produce movements along the same two degrees of freedom, and that there was a straightforward mapping of these two directions between the EMG space and the formant space. This finding suggests that there could be an acoustic basis, related to the perception of speech, for the degrees of freedom of the tongue movements and that speakers could learn the synergistic muscle actions that they use to produce the desired acoustics. This hypothesis is supported by the observation that in phonetic space, the direction of the first degree of freedom corresponds to a movement along a natural vowel axis, /i , e, ɛ, æ , a/; however, this is not the case for the second degree of freedom. Furthermore, to produce the basic high-low and front-back distinctions, it is necessary to use the two degrees of freedom in combination. Therefore, at least the second of these main degrees of freedom does not seem to be a function of basic phonetic categories.

In this paper we explore the hypothesis that these two main degrees of freedom could find their origins in the anatomical and biomechanical properties of the speech production apparatus. Toward this aim, a two-dimensional, physiologically-based model of the tongue (Payan and Perrier, 1997 ; Perrier et al., 1998) was used to generate a large set of tongue configurations, on which a PCA was run in order to extract the main axes of deformation.

## 2. THE 2D TONGUE MODEL

### 2.1. Biomechanical structure

The tongue model (an improved version of the model of Payan and Perrier, 1997) includes the main muscles responsible for shaping and moving the tongue in the midsagittal plane: posterior and anterior parts of the genioglossus (GGP and GGA), styloglossus (STY), hyoglossus (HYO), inferior and superior longitudinalis (IL and SL) and verticalis (VER). Elastic properties of tissues are accounted for by finite-element (FE) modeling of the tongue mesh in 2D defined by 221 nodes and 192 hexahedric elements. Muscles are modeled as force generators that (1) act on anatomically specified sets of nodes of the FE structure, and (2) modify the stiffness of specific elements of the model to account for muscle contractions within tongue tissues. Curves representing the contours of the lips, palate and pharynx in the midsagittal plane are added. The jaw and the hyoid bone are represented in this plane by static rigid structures to which the tongue is attached. Changes in jaw height can be simulated through a single parameter that modifies the vertical position of the whole FE structure in relation to the palate.

### 2.2. Control of the model

The model is controlled according to Feldman's Equilibrium Point Hypothesis (Feldman, 1986). This theory of motor control, grounded in basic neurophysiological mechanisms of muscle force generation, suggests that the central nervous system controls movements by selecting, for each acting muscle, a threshold muscle length, $\lambda$, where the recruitment of $\alpha$ motoneurons (responsible for active forces) starts. If the muscle length is larger than $\lambda$, muscle force increases exponentially with the difference between the two lengths. Otherwise no active muscle force is generated. The muscle control space is thus called the $\lambda$ space. Moreover, Feldman's basic suggestion is that movements are produced from posture to posture, a posture being a stable mechanical equilibrium state of the motor system associated to a specific set of $\lambda$ values. Hence, in the model, a sequence of discrete control variable values ($\lambda$s), those specifying the successive postures, underlies a continuous movement. In the current version of our control model, the transition from one $\lambda$ set to the next is made with constant rate $\lambda$ shifts, and the onset and offset times of the $\lambda$ shifts are the same for all muscles. In other words, we hypothesize that the recruitments of all tongue muscles are synchronized.

## 3. PRINCIPAL COMPONENT ANALYSIS

As already mentioned in the introduction, the first attempt to use a biomechanical model of the speech production apparatus to extract the degrees of freedom of speech articulations was provided by Sanguineti and colleagues (Sanguineti et al., 1997; Sanguineti et al., 1998). Their approach was fundamentally data driven: starting from Xray views of French articulations, these authors inferred, according to certain constraints, the control vectors in the $\lambda$ space, and then they computed a PCA on the distribution of the control vectors. The advantage of this approach is that the link between the degrees of freedom and the muscle commands are completely known. However, since the data were collected during speech production, this approach does not permit a study of the more general influences of the anatomy and biomechanics on tongue deformation.

In our study, the tongue model is used to generate a large number of tongue shapes, and we ran a PCA on the resulting sets of geometrical configurations. Therefore, our study involves no speech-specific control. From this perspective, our approach is thus more general, and should provide a better view of the influence of the anatomy and biomechanics on tongue deformation. However, the drawback is that the link between the extracted degrees of freedom and the muscle variables is far from obvious.

### 3.1. Generating the data set

The generation of the set of tongue shapes was made using a classical Monte-Carlo method. First we define rest position as the position of the tongue in which no active muscle force is generated. In the rest position the passive elasticity forces, internal to the tongue, are just balanced by the force of gravity. The commands at rest were then determined so that no active muscle force is generated in this position, and also that muscle force begins to be generated as soon as the tongue is shifted, even slightly, from its rest position. Second, to specify the sampling of the control space, the assumption was made that, for each muscle, the distribution of values of the control variable around its mean value is Gaussian. The standard deviation of the Gaussian distribution was determined in the following manner. For each muscle, a small value and a large value of $\lambda$, corresponding to -2 and +2 standard deviations from the mean value, respectively, were determined as follows. The small $\lambda$ values were chosen to be values smaller than the rest value that induced, for each muscle, a maximum shift of the tongue contour from the rest position of 15 mm for the large muscles (GGP, HYO, STY, GGA), and of 8 mm for the small muscles (VER, IL, SL). The large lambda values were fixed to be equal to the rest lambda values plus 15mm (for large muscles) or 8mm (for small muscles).

The control space was then sampled according to this distribution, resulting in the generation of 1800 tongue shapes with the model. This number was chosen because it is high enough to provide a stable statistical analysis. The jaw-height parameter was kept constant

Figure 1 shows the resulting distributions of the 17 nodes on the dorsal tongue contour. Dispersion ellipses at 2 standard deviations are superimposed on the distributions. It can be seen that a large range of variation is provided, and that no consonantal configuration is observed. Avoiding consonantal configurations is important, since contacts with the vocal tract walls would induce tongue deformations having their origins external to the tongue.

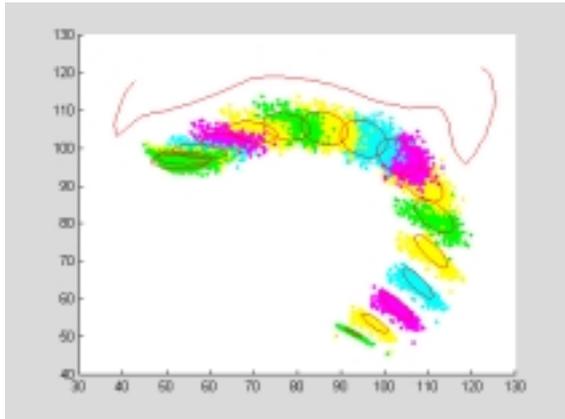

**Figure 1:** Distributions of the positions of the 17 nodes on the upper contour of the tongue. Dispersion ellipses correspond to 2 standard deviations. The upper line corresponds to the palate contour. Teeth are on the left, the velum is on the right. Distances in X and Y are in mm.

## 3.2. Extracting the main factors

A principal component analysis was run on the 17 sets of node positions. For the analysis, the data were first normalized: the average values were computed for each set and they were subtracted from the original X and Y values; the obtained X and Y values were then divided by their standard deviations. We thus obtained a set of 17*1800 normalized values for X and Y. The correlation matrix was then computed, and the eigenvectors gave the directions of the degrees of freedom. The larger the eigenvalue, the more important the contribution of the degree of freedom to the tongue deformation.

Four factors accounted for 97.4 % of the variance of the data. The impacts of the first three factors (referred to as F1, F2, and F3) on the tongue deformation are represented in Figures 2 to 4, for a shift along each axis of +2 and –2 standard deviations. The first factor accounts for 58.3% of the variance, the first two factors for 84.6% and the first three factors for 94.2%

## 3.3. Discussion

These three factors are very similar to the ones found for English by Nix et al. (1996) (see Fig 4 p 3713). In particular, the first factor effectively corresponds to the *front raising* factor, while the second factor is comparable to the *back raising* factor.

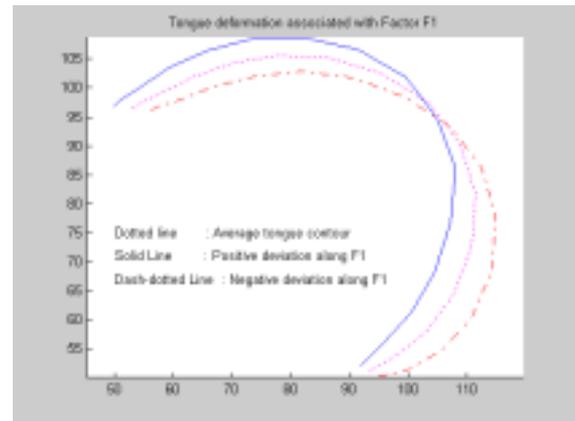

**Figure 2:** Effect of the first factor on the tongue shape

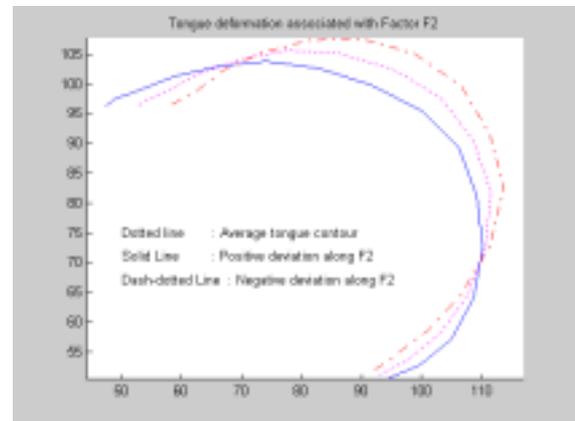

**Figure 3:** Effect of the second factor on the tongue shape

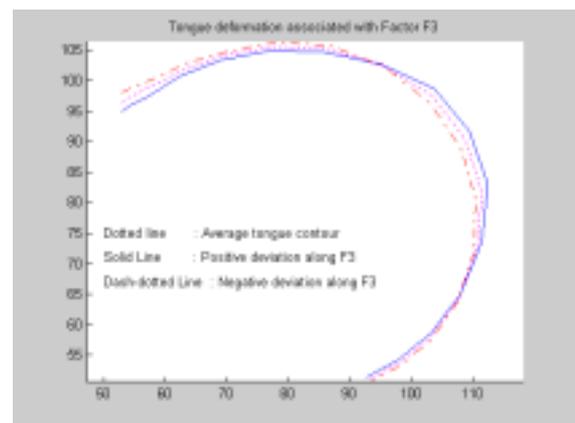

**Figure 4:** Effect of the third factor on the tongue shape

It is also interesting to observe that, in the majority of the studies based on statistical analyses of articulatory data, more than 90% of the variance observed for a subject were described by the first two factors, while in our study three factors are necessary to reach the same level of description. This difference is in agreement with the Nix et al. (1996) findings, which showed that when the tongue shapes of 6 speakers were analyzed together, 4 factors were necessary to reach the same level of description as 2 factors extracted from the data of a single subject. Since our data were generated from a model, they may be more general, analogous to the combined data from 6 speakers.

## 4. CONTROL VARIABLES AND DEGREES OF FREEDOM

Finding the main degrees of freedom of the tongue deformation is an important first step toward the elaboration of a complete model of control of speech production. However, if we want to understand how the muscles are controlled in order to move the tongue in the space of its degrees of freedom, it is necessary to provide a link between the muscle control variables and the degrees of freedom. Some general information about this link can be found in our data. This is illustrated by Figure 5: the antagonist pairs GGP/HYO and GGA/STY found by Maeda and Honda (1994) are associated with the control of the tongue deformation along the first and second degree of freedom, respectively. But at the same time, it shows the complexity of the search for a proper mathematical formalization of the link between the functional and the muscle control spaces. Indeed, while the relations look quite linear for small $\lambda$ values, they become more noisy for large values. This can be easily explained. When $\lambda$ is small, the muscle is essentially active regardless of the tongue shape; therefore, changes in $\lambda$ induce changes in tongue shape. However, given the muscle force generation mechanism proposed in the Equilibrium-Point hypothesis, this is no longer the case for large $\lambda$ values, where the muscle is not active for a majority of the tongue shapes. Hence, the relationship between the $\lambda$ commands and the movement in the space of the degrees of freedom is highly non-linear. The use of non-linear mapping techniques should permit a better understanding of the synergetic and antagonistic relations among muscles associated with the control of speech movements in the degrees of freedom space.

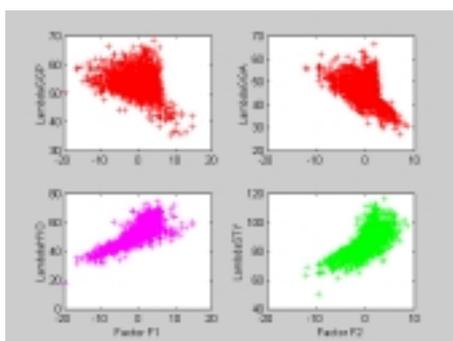

**Figure 5:** First factor F1 *versus* GGP / HYO commands (left); Second factor F2 *versus* GGA / STY commands (right).

## 5. CONCLUSION

Our study suggests that the degrees of freedom extracted for different languages from articulatory data are not speech-specific, but are due to the anatomical and biomechanical properties of the tongue. Speech control would then use these degrees of freedom to determine and differentiate the articulations of the different sounds of a language. The way speakers control their muscles in relation to the degrees of freedom is under investigation using non-linear mapping techniques.

**Acknowledgements:** This research was supported by the CNRS (France), the NSF and NIH (USA)